\documentclass[journal]{IEEEtran}
\usepackage{color}
\usepackage{multirow}
\usepackage{cite}
\usepackage{balance}
\usepackage{amsmath}
\usepackage{enumitem}
\usepackage[]{algorithm2e}
\usepackage{url}
\setlist[itemize]{noitemsep, topsep=0pt}

\usepackage{pdfsync}

\ifCLASSINFOpdf
\usepackage[pdftex]{graphicx}
\DeclareGraphicsExtensions{.pdf,.jpeg,.png}
\else
\usepackage[dvips]{graphicx}

\DeclareGraphicsExtensions{.eps}
\fi
\graphicspath{{figs/}}
\usepackage{epstopdf}
\usepackage{booktabs}
\usepackage{amsmath}

\begin{document}

    \title{Transmission-Distribution Co-Simulation: Analytical Methods for Iterative Coupling}
    %
    %
    %
    
    \author{Gayathri Krishnamoorthy,~\IEEEmembership{Student Member,~IEEE,}
        and Anamika Dubey,~\IEEEmembership{Member,~IEEE}
        \thanks{G. Krishnamoorthy and A. Dubey  are with the School of Electrical Engineering and Computer Science, Washington State University, Pullman, WA, 99164 e-mail: g.krishnamoorthy@wsu.edu, anamika.dubey@wsu.edu.}
    }
    \maketitle

\begin{abstract}
With the increased penetrations of distributed energy resources (DERs), the need for integrated transmission and distribution system analysis (T\&D) is imperative. This paper presents an integrated unbalanced T\&D analysis framework using an iteratively coupled co-simulation approach. The unbalanced T\&D systems are solved separately using dedicated solvers. An iterative approach is developed for T\&D interface coupling and to ensure convergence of the boundary variables. To do so, analytical expressions governing the T\&D interface are obtained. First-order and second-order convergent methods using Fixed-point iteration (FPI) method and Newton's method, respectively are proposed to solve the system of nonlinear T\&D interface equations. The proposed framework is tested using an integrated T\&D system model comprised of 9-bus IEEE transmission test system integrated with a real-world 6000-bus distribution test system. The results show that the proposed framework can model the impacts of system unbalance and increased demand variability on integrated T\&D systems and converges during stressed system conditions. As expected, Newton's method converges faster with a fewer number of iterations as compared to FPI method and the improvements are more pronounced during high levels of system unbalance and high loading conditions.          
        
\end{abstract}
    
    
    %
    \IEEEpeerreviewmaketitle
        \vspace{-0.3 cm}
%
%
    \section{Introduction}    
    \IEEEPARstart{W}ith the incentivized rapid decarbonization of electric power generation industry and aggressive renewable portfolio standards (RPS), the electric power delivery (T\&D) system is expected to transform rapidly in the foreseeable future \cite{USRPSReport}. Several exploratory studies have pointed out that the ongoing and future large-scale DER deployment projects can potentially affect the regional transmission grid operations \cite{evans2014new,palmintier2016experiences}. 
In order to evaluate and mitigate the associated challenges, new tools capable of capturing the interactions between T\&D systems are required. The inherent differences in the modeling and analysis of the two systems coupled with the problem of solving a very large-scale integrated T\&D system model makes this a non-trivial problem \cite{palensky2017cosimulation}.

A majority of the existing DER interconnection studies utilize a decoupled T\&D system model and ignore T\&D interactions. Owing to the need for capturing T\&D interactions, lately, increasing efforts have been put forward to develop simulation tools capable of combined T\&D modeling and analysis. The existing frameworks can be broadly categorized as 1) standalone T\&D system models \cite{evans2010verification,chassin2014gridlab}, and 2) co-simulation platforms \cite{ciraci2014fncs,palmintier2017igms, palmintier2017design, balasubramaniam2017combined,huang2017integrated, sun2015master,huang2017comparative}. The standalone T\&D models are computationally intensive and not practical for simulating a large-scale integrated T\&D system \cite{palmintier2017igms}. On the other hand, a co-simulation platform efficiently simulates a large-scale integrated T\&D system. A T\&D co-simulation framework interfaces the simulators of multiple interacting domains by enabling communication, data exchange, and time synchronization. Owing to their computational advantage, several frameworks including FNCS \cite{ciraci2014fncs}, IGMS \cite{palmintier2017igms}, HELICS \cite{palmintier2017design} have been recently developed to facilitate T\&D co-simulation.

The existing T\&D co-simulation frameworks, however, are limited in accurately modeling T\&D interactions as these conduct transmission system analyses using a balanced positive sequence power flow that does not account for unbalanced load conditions \cite{ciraci2014fncs,palmintier2017igms, palmintier2017design, balasubramaniam2017combined}. Furthermore, the existing co-simulation methods loosely couple T\&D systems introducing errors in the solutions \cite{ciraci2014fncs,palmintier2017igms}. A few recent articles address this problem. For example, in \cite{balasubramaniam2017combined}, a tightly coupled interface protocol was used for quasi-static T\&D simulations. However, the transmission network was solved using a positive sequence model that cannot capture unbalanced loading conditions. 
In our prior work, an iteratively coupled co-simulation framework for unbalanced T\&D systems was developed, however, the convergence was achieved by simply exchanging the boundary variables \cite{DubeyPES2018}. Similarly, in \cite{huang2017integrated}, authors have solved an unbalanced three-phase integrated T\&D system model where boundary variables were exchanged iteratively. The existing literature including our prior work does not provide a mathematical analysis of the iterative co-simulation interface. Furthermore, the rate of convergence for iterative co-simulation methods for a large-scale system will increase with the adverse load conditions requiring new methods to enable a faster convergence.


In this paper, an approach to iteratively couple T\&D models using co-simulation is developed that is suitable for the integrated analysis of T\&D systems subject to unbalanced load conditions and significant variations in load demand. The proposed framework is comprised of three modules: a three-sequence AC power flow for the transmission system, a three-phase AC power flow for the distribution system, and an iterative coupling approach at the T\&D interface. 
The novelty of the work lies in the proposed analytical formulations for the boundary variable update rules for iterative T\&D coupling also termed as co-iteration rules. 
To this regard, we obtain models for T\&D co-simulation interface and propose first-order and second-order convergent techniques based on Fixed-point iteration (FPI) and Newton's method to solve the associated nonlinear T\&D interface equations. The proposed iterative coupling technique can be easily incorporated into any existing co-simulation platform capable of co-iteration such as HELICS. It should be noted that this work does not intend to replace the existing large-scale co-simulation platforms, but aims to propose advances that will help refine the existing modeling and co-simulation techniques. The major focus is on developing improved methods with faster convergence for iterative coupling of accurately modeled T\&D systems.


Both first- and second-order methods proposed to solve iteratively coupled T\&D system are validated for stressed system conditions. The test system is comprised of a 9-bus IEEE transmission test system \cite{TransTestFeeder} integrated with EPRI Ckt24 distribution feeders \cite{DistTestFeeder}. It is observed that Newton's method converges faster when compared to FPI method and the computational advantages are more pronounced for severe unbalance and high loading conditions. It should be noted that this work focuses on developing T\&D interface for quasi-static power system co-simulation. The proposed update rules can potentially be extended to iteratively coupled dynamic T\&D co-simulation. However, the analysis of dynamic co-simulation is not within the scope of this work. The major contributions of this paper are as follows:    
    \begin{enumerate} [noitemsep,topsep=0pt,leftmargin=*]
        \item \textit{Mathematical representation of T\&D co-simulation interface} - The co-simulation interface is represented using a set of nonlinear equations that appropriately represent the interface coupling and individual subsystem equations.
        \item \textit{First-order and second-order updates using Fixed-point iteration and Newton's method} - The nonlinear interface equations are solved iteratively. The first-order update employs FPI algorithm while second-order updates are based on Newton's method using Jacobian-based update rule.
        \item \textit{Accurate simulation during system unbalance and demand variability} - The strength of the proposed iterative coupling framework in modeling the demand unbalance and variability issues is demonstrated using multiple case studies. Also, it is validated that the three-sequence transmission model leads to an accurate T\&D co-simulation during unbalance.
    \end{enumerate}

\textcolor{blue}{The rest of the paper is organized as follows. Section II presents the need for the tightly coupled T\&D co-simulation platform. Section III describes the proposed co-simulation approach. Section IV presents the mathematical model for co-simulation and derives update rules for FPI and Newton's method. Section V details results and discussions followed by conclusion in Section VI.}

\section{Need for Iteratively Coupled T\&D Co-simulation Framework}\textcolor{blue}{
As discussed previously, the combined T\&D simulation can be achieved using 1) Stand-alone T\&D system models and 2) Co-simulation approach. The major limitation of the standalone unified modeling approach is the cost of simulation. Given that the detailed model of a typical distribution feeder includes 1000s of buses/nodes, a stand-alone T\&D model is usually too complex to simulate and analyze using a single tool. In addition, the stand-alone models do not take the advantage of legacy power systems modeling and simulation platform. It should be noted that the electric power transmission systems and distribution systems are significantly different. While transmission systems are largely balanced with low R/X ratio and highly meshed, a typical distribution feeder is highly unbalanced, include single-phase loads and laterals, and is radial in configuration. Owing to these differences, the solution approach used for the two networks also differ. The Newton-Raphson method is adopted to solve power flow model for transmission systems while distribution systems are solved using forward-backward sweep or current injection methods. There are multiple other functional differences between the two systems making it impractical and inefficient to bring together all of the functionalities of individual legacy software tools into a single simulation environment. Consequently, it is more efficient to use co-simulation methods that bring the individual legacy tools together to perform the combined T\&D simulation studies without having to make changes to the individual legacy platforms.} 

\textcolor{blue}{Next, we discuss the need for an iteratively/tightly coupled T\&D co-simulation approach. The loosely coupled co-simulation methods are accurate only when the changes in distribution system loading characteristics, both load unbalance and demand variability, are slower than the simulation time-step at which the two systems are solved and the solutions are exchanged. Otherwise, the loosely coupled model introduces simulation errors \cite{palmintier2017igms}. This is because, in the loosely coupled models, the time step for individual T\&D simulators is advanced without making the boundary variables converge. The primary assumption is that the changes in power system loads are rather slow and the system converges over multiple time steps. This limits the applicability of the existing framework when modeling faster load/DER variations. In an actual co-simulation platform, the simulation time step must not advance until the boundary variables for both transmission and distribution systems have converged. This requires an iteratively or tightly coupled co-simulation approach.} 


\vspace{-0.2cm}

\section{Transmission and Distribution Co-simulation Framework}
The primary objective of this paper is to bring the co-simulation platform closer to the standalone unified models of T\&D systems. To do so, transmission and distribution systems are modeled in full three-phase representation. For co-simulation, a novel iterative framework is proposed that allows for T\&D coupling by iteratively solving nonlinear T\&D interface equations using first-order and second-order convergent methods. 

\vspace{-0.3cm}

\subsection{Transmission and Distribution System Quasi-Static Model}
The co-simulation platforms currently available in the literature, perform transmission system analysis using a balanced positive sequence AC power flow. The increase in DER penetrations on distribution systems may increase demand unbalance rendering the positive sequence models inapplicable for integrated T\&D system analysis. Therefore, a detailed three-phase power flow analysis is required for transmission systems.

In this paper, a three-sequence model for the transmission system is developed using a sequence component method \cite{abdel2005improved}. The three-sequence model can represent the effects of unbalanced loads and untransposed transmission lines while not significantly increasing the computational complexity of the power flow analysis. The three-phase transmission system is decoupled into three independent sequence circuits by replacing the off-diagonal elements with the respective compensation current injections \cite{abdel2005improved}. The decoupled three-sequence models are solved separately. The positive sequence model is solved using the Newton-Raphson method. The negative and zero sequence models are solved using linear equations. The process is iterative and repeated until the change in positive sequence power flow due to negative and zero sequence components is within the tolerance. Please refer to \cite{abdel2005improved} for details. Similarly, the distribution system is modeled in full three-phase representation. The three-phase modeling and quasi-static analysis is done using OpenDSS \cite{OpenDSS}.

\begin{figure}[t]
        \centering
        \includegraphics[width=0.5\textwidth]{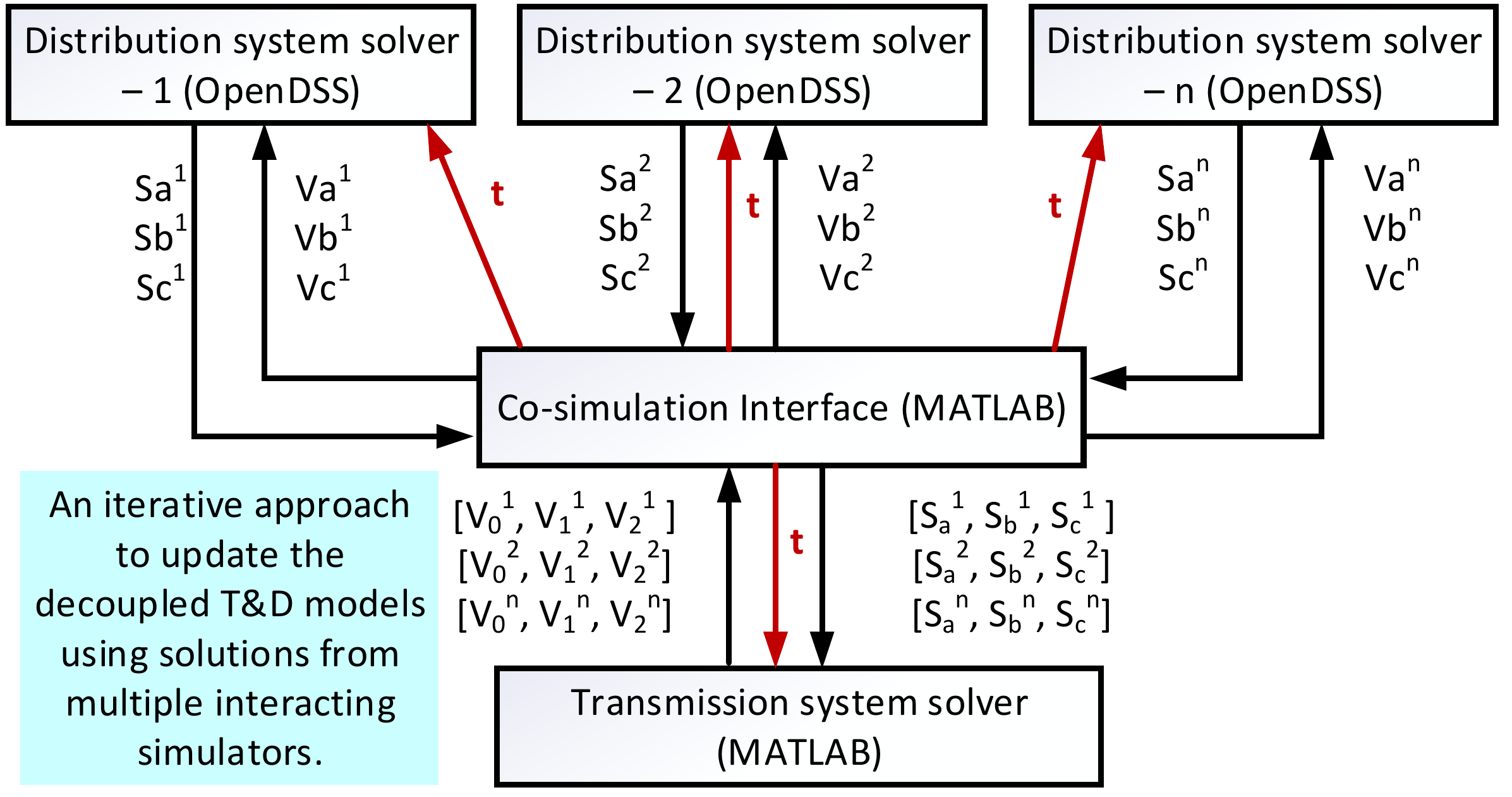}
\vspace{-0.6cm}
        \caption{Proposed T\&D Co-simulation interface.}
        \label{fig1}
\vspace{-0.6cm}
    \end{figure}

\vspace{-0.3cm}

\vspace{-0.1cm}
\subsection{Co-simulation Approach}
The main contribution of this paper is to develop faster algorithms to iteratively couple the T\&D systems and allow for interface convergence using multiple co-simulation iterations/co-iterations. The proposed co-iteration approach analytically solves the nonlinear T\&D interface equations using first-order and second-order convergent methods and provides a mathematical treatment to the co-simulation problem. The general principle is as follows. At a given time step of the quasi-static simulation, the T\&D systems are solved separately using their respective solvers. This step solves decoupled T\&D models where the transmission system solver models the connected distribution network as an equivalent load and the distribution solver models the upstream transmission bus as a voltage source. The bus voltages and angles obtained from transmission network solver and active and reactive power flow obtained from distribution network solver at the point-of-common-coupling (PCC) are referred to as boundary variables (see Fig. 1). After decoupled T\&D systems are solved, boundary variables are tested for convergence. If the tolerance limit is not satisfied, the co-simulation stage begins. At a given co-simulation iteration (co-iteration), T\&D systems have partially correct information about the boundary variables that need updating. The decoupled representations for T\&D networks are updated at each co-iteration using the proposed update rules for boundary variables, further detailed in Section III. With updated boundary variables, the decoupled T\&D models are solved again. The co-iterations are repeated until the errors in boundary variables obtained from decoupled models are within the pre-specified tolerance. 

\vspace{-0.3cm}

\subsection{Interface Architecture and Time Coordination}
The transmission system modeling and three-sequence power flow analysis is done using MATLAB. The modeling and analysis of the three-phase distribution system is done using OpenDSS. The interface protocol for T\&D simulators and their time synchronization is done using a master algorithm written in MATLAB, referred to as co-simulation interface (see Fig. 1). The module for co-simulation interface synchronizes the time between different simulators and implements the co-iteration protocols. The co-simulation interface provides two types of output to all interacting simulators: a timing signal and boundary variable updates. The timing signal ensures that none of the simulators advance to the next time step until the integrated T\&D model has converged for the current time-step. The co-simulation interface also includes an algorithm for co-iteration that updates the boundary variables and uses internal logic to evaluate if the convergence has reached and should the simulation advance to the next time step. Note that this paper employs a parallel or Jacobi communication scheme \cite{huang2017comparative}. This means that the variables are simultaneously updated for all interacting simulators. 

\vspace{-0.2cm}
\section{Analytical Model for T\&D Interface and Iterative Coupling/Co-iteration}
An analytical model is developed for the abstract representation of the coupled T\&D system using a set of nonlinear equations for subsystem solvers and co-simulation interface. First-order and second-order convergent methods are proposed to solve the nonlinear interface equations.
\vspace{-0.2cm}

\subsection{Interface Equations}
In quasi-static model for coupled T\&D system, transmission system is solved using three-sequence analysis while distribution system is solved using a three-phase power flow. The input and output parameters for the subsystem representing transmission system are complex per-phase power demand at the PCC ($S_T$) and sequence voltages at PCC ($V_T$), respectively. Similarly, the input and output parameters for the subsystem representing distribution system are phase voltages at PCC ($V_D$), and complex per-phase power demand at the PCC ($S_D$), respectively. The mathematical equations governing the coupled system are given in (1)-(4).

\vspace{-0.3cm}
\begin{small}
\begin{eqnarray}
  V_T &=& f_1(S_T) \\
  S_D &=& f_2(V_D) \\
  S_T - S_D &=& 0 \\
  \mathcal{T}(V_D) - V_T &=& 0
\end{eqnarray}
\end{small}
\noindent where,\\
$f_1(x)$ - nonlinear equations for three-sequence power flow.\\
$f_2(x)$ - nonlinear equations for three-phase power flow. \\
\begin{small}
$S_T = \left[
         \begin{array}{c}
           S_a \\
           S_b \\
           S_c \\
         \end{array}
       \right]$,
$S_D = \left[
         \begin{array}{c}
           S_a \\
           S_b \\
           S_c \\
         \end{array}
       \right]$,
$V_T = \left[
         \begin{array}{c}
           V_0 \\
           V_1 \\
           V_2 \\
         \end{array}
       \right]$,
$V_D = \left[
         \begin{array}{c}
           V_a \\
           V_b \\
           V_c \\
         \end{array}
       \right]$,
       $\mathcal{T}  = \dfrac{1}{3}\left[
   \begin{array}{ccc}
     1 & 1 & 1 \\
     1 & a & a^2 \\
     1 & a^2 & a \\
   \end{array}
 \right]$.
 \end{small}
\\

Note that (1) and (2) are solved using the subsystem solvers. The mathematical model for the co-simulation interface is detailed next. We define interface equations as a set of non-linear equations in (5) and (6).

\vspace{-0.3cm}
\begin{small}
\begin{eqnarray}
I_T(S_T,V_D) &:& S_T - f_2(V_D) =0 \\
I_D(S_T,V_D) &:& \mathcal{T}(V_D) - f_1(S_T) = 0
\end{eqnarray}
\end{small}

\vspace{-0.5cm}

The residual components for each system based on a given subsystem solution are defined as the following:

\vspace{-0.5cm}
\begin{small}
\begin{eqnarray}
R_T &=& S_T - f_2(V_D)  \\
R_D &=& \mathcal{T}(V_D) - f_1(S_T)
\end{eqnarray}
\end{small}

\vspace{-0.5cm}

The co-simulation interface shown in Fig. 1 solves non-linear equations defined in (5) and (6) subject to (1) and (2). The approach is detailed in Algorithm 1. First, subsystem equations, (1) and (2), are solved in parallel. The roots (output) of the subsystem equations are used for residual evaluation of the coupled system at the interface using (7) and (8). A global interface residual vector ($\mathcal{R}$) is defined to evaluate the condition for the convergence of the co-simulation framework, where $\epsilon_1$ and $\epsilon_2$ are predefined tolerance parameters (9). The objective is to iteratively solve interface equations defined in (5) and (6) until the residual evaluated using (7) and (8) are within a permissible error tolerance. If convergence criteria is not met, the boundary variables are updated. The update rules for boundary variables are derived in sections III-B and III-C for FPI and Newton's method, respectively. The process is repeated until the boundary variables converge.
\begin{equation}
\small
\mathcal{R}=
\left[
\begin{array}{c}
R_T\\
R_D\\
\end{array}
\right] \leq \left[
\begin{array}{c}
\epsilon_1\\
\epsilon_2\\
\end{array}
\right]
\end{equation}

\vspace{-0.4cm}

\subsection{Fixed-Point Co-iteration Method}
The FPI method is one of the classic Jacobian-free solution techniques to solve nonlinear system of equations. The approach is to transform the root finding problem to a fixed-point problem. For a set of non-linear equations defined as $f(x) = 0$, the FPI update sequence is given as the following (10), where $\alpha \in \mathcal{R}$ is relaxation parameter:
\begin{equation}
\small
x_{n+1}=x_n \pm \alpha f(x_n)
\end{equation}

Using FPI method, the updates for solving (5)-(6) are defined in (11).
\begin{equation}
\small
\left[
\begin{array}{c}
S_T(n+1)\\
V_D(n+1)\\
\end{array}
\right] =
\left[
\begin{array}{c}
S_T(n)\\
V_D(n)\\
\end{array}
\right] + \alpha
\left[
\begin{array}{c}
S_T(n) - f_2(V_D(n)) \\
V_D(n) - \mathcal{T}^{-1}(f_1(S_T(n)))
\end{array}
\right] \\
\end{equation}

For $\alpha = -1$, the FPI updates are given by:
\begin{equation}
\small
\left[
\begin{array}{c}
S_T(n+1)\\
V_D(n+1)\\
\end{array}
\right] =
\left[
\begin{array}{c}
f_2(V_D(n))\\
\mathcal{T}^{-1}(f_1(S_T(n)))\\
\end{array}
\right] \\
\end{equation}

\begin{algorithm}[t]
\small
    \mbox{\emph{Algorithm 1: Solving Coupled System: FPI and Newton's method}}\\
    \noindent\rule{8.4cm}{1pt}\\
    {Initialize time index for the time-series simulation, t = 1.}\\
    \For {t=1:$t_{step}$:T}    {
    {Initialize Input Variables: ${_t}S_T$, ${_t}V_D$}\\
    {Initialize iteration count, n = 1} \\
    {Solve subsystems in parallel} \\
    ${_t}V_T(n)= f_1({_t}S_T(n)) $\\
          ${_t}S_D(n)= f_2({_t}V_D(n)) $\\
    \vspace{3pt}
    {Check residual at the interface} \\
    \vspace{3pt}
    ${_t}\mathcal{R}(n)=
    \left[
    \begin{array}{c}
    {_t}S_T(n) - {_t}S_D(n)\\
    {_t}V_D(n) - \mathcal{T}^{-1}({_t}V_T(n))
    \end{array}
    \right]$    \\
          //   Iteration loop     \\        
        \While {$|{_t}\mathcal{R}(n)| \geq \epsilon$}{
        {Update boundary variables for next iteration} \\
    {Update rule for FPI method}\\
\vspace{3pt}
            $\left[
            \begin{array}{c}
        {_t}S_T(n+1)\\
        {_t}V_D(n+1)\\
            \end{array}
            \right]$ =
            $\left[
            \begin{array}{c}
            {_t}S_D(n)\\
            \mathcal{T}^{-1}({_t}V_T(n))\\
            \end{array}
            \right] $\\
\vspace{3pt}
        {Update rule for Newton's Method}\\
\vspace{3pt}
            $\left[
                \begin{array}{c}
                {_t}S_T(n+1)\\
                {_t}V_D(n+1)\\
                \end{array}
                \right]$ =
                $\left[
                \begin{array}{c}
                {_t}S_T(n)\\
                {_t}V_D(n)\\
                \end{array}
                \right]$ - $\left[
                \begin{array}{c}
                \Delta S_T\\
                \Delta V_D\\
                \end{array}
                \right]$ \\
\vspace{3pt}
        {Increment iteration count}\\
        n = n+1 \\
        {Solve subsystems in parallel} \\
        ${_t}V_T(n)= f_1({_t}S_T(n)) $\\
          ${_t}S_D(n)= f_2({_t}V_D(n)) $\\
     \vspace{3pt}
     {Check residual at the interface} \\
     \vspace{3pt}
    ${_t}\mathcal{R}(n)=
    \left[
    \begin{array}{c}
    {_t}S_T(n) - {_t}S_D(n)\\
    {_t}V_D(n) - \mathcal{T}^{-1}({_t}V_T(n))
    \end{array}
    \right]$    \\
    }    
}
    \noindent\rule{8.4cm}{1pt}\\
    \label{algo}
    \vspace{-20pt}
\end{algorithm}

The algorithm for implementing the FPI method for the coupled T\&D system is detailed in Algorithm 1. The co-simulation interface only focuses on solving interface equations and updating input to the two subsystems i.e. $S_T$ and $V_D$. The nonlinear equations for individual subsystems are solved using subsystem solvers. Until the convergence criteria is met, the input to the individual subsystems is updated using the FPI co-iteration sequence defined in (11). For $\alpha = -1$, the FPI updates are the subsystem solutions obtained in the previous time step. This method, therefore, iteratively exchanges the subsystem solutions until the boundary variables converge.

\vspace{-0.4cm}

\subsection{Newton-based Co-iteration Method}
The Newton's method to solve the non-linear system of equations, $f(x)=0$, requires the first derivative of the function i.e. $\mathcal{J}(f(x))$. The iterative sequence for solving $f(x)=0$ using Newton's method is given in (13).
\begin{eqnarray}
\nonumber x_{n+1}&=&x_n - \Delta{x}\\
\Delta{x} &=& \mathcal{J} (f(x_n))^{-1}  f(x_n)
\end{eqnarray}
where, $\mathcal{J}$ is the Jacobian operator.
%
%

For the T\&D co-simulation problem, the system of nonlinear equations to be solved are the interface equations $I_T(S_T, V_D)$ and $I_D(S_T, V_D)$ defined in (5)-(6). The algorithm for implementing Newton's method for the coupled T\&D system is detailed in Algorithm 1 and described as follows. Same as FPI method, the interface equations are solved iteratively until the residuals defined in (7)-(8) are within the permissible error tolerance. Until the convergence criteria is met, the input to the individual subsystems, i.e. $S_T$ and $V_D$, are updated using Jacobian-based update rule derived in the following sections.

\subsubsection{Defining Jacobian Operator and Update Rule}
This section presents the jacobian-based update rule for solving (5) and (6). First, we define jacobian operator $\mathcal{J}$ obtained by differentiating (5) and (6) wrt. variables $S_T$ and $V_D$.
\begin{equation}\label{Jac}
\small
\mathcal{J} = \left[\begin{array}{cc}
                \dfrac{\partial I_T}{\partial S_T} & \dfrac{\partial I_T}{\partial V_D} \\
                \dfrac{\partial I_D}{\partial S_T} & \dfrac{\partial I_D}{\partial V_D}
              \end{array}\right]= \left[\begin{array}{cc}
                                     \dfrac{\partial S_T}{\partial S_T} & -\dfrac{\partial f_2(V_D)}{\partial V_D} \\
                                     -\dfrac{\partial f_1(S_T)}{\partial S_T} & \dfrac{\partial \mathcal{T} V_D}{\partial V_D}
                                   \end{array}\right]
\end{equation}
where,
\begin{equation}
\small
\nonumber
\dfrac{\partial S_T}{\partial S_T} = \left[\begin{array}{cc}
    \dfrac{\partial P_T}{\partial P_T} & \dfrac{\partial P_T}{\partial Q_T} \\
    \dfrac{\partial Q_T}{\partial P_T} & \dfrac{\partial Q_T}{\partial Q_T}
  \end{array}\right] = \left[\begin{array}{cc}
    I & 0 \\
    0 & I
  \end{array}\right], I = \left[\begin{array}{ccc}
                            1 & 0 & 0 \\
                            0 & 1 & 0 \\
                            0 & 0 & 1
                          \end{array}\right]
\end{equation}
\begin{equation}
\small
\nonumber  \dfrac{\partial f_2(V_D)}{\partial V_D} =
  \left[\begin{array}{cccccc}
    \dfrac{\partial P_a}{\partial |V_a|} & \dfrac{\partial P_a}{\partial |V_b|} & \dfrac{\partial P_a}{\partial |V_c|} & 0 &0 &0\\
    \dfrac{\partial P_b}{\partial |V_a|} & \dfrac{\partial P_b}{\partial |V_b|} & \dfrac{\partial P_b}{\partial |V_c|} & 0 &0 &0\\
    \dfrac{\partial P_c}{\partial |V_a|} & \dfrac{\partial P_c}{\partial |V_b|} & \dfrac{\partial P_c}{\partial |V_c|} & 0 &0 &0\\
    0 &0 &0 & \dfrac{\partial Q_a}{\partial |V_a|} & \dfrac{\partial Q_a}{\partial |V_b|} & \dfrac{\partial Q_a}{\partial |V_c|} \\
    0 &0 &0 & \dfrac{\partial Q_b}{\partial |V_a|} & \dfrac{\partial Q_b}{\partial |V_b|} & \dfrac{\partial Q_b}{\partial |V_c|} \\
    0 &0 &0 & \dfrac{\partial Q_c}{\partial |V_a|} & \dfrac{\partial Q_c}{\partial |V_b|} & \dfrac{\partial Q_c}{\partial |V_c|}
  \end{array}\right]
\end{equation}
\begin{equation}
\small
\nonumber \dfrac{\partial \mathcal{T}V_D}{\partial V_D} = \mathcal{T},
\dfrac{\partial f_1(S_T)}{\partial S_T} = \left[
\begin{array}{ccc}
  \dfrac{\partial V_0}{\partial S_a} & \dfrac{\partial V_0}{\partial S_b} & \dfrac{\partial V_0}{\partial S_c} \\
  \dfrac{\partial V_1}{\partial S_a} & \dfrac{\partial V_1}{\partial S_b} & \dfrac{\partial V_1}{\partial S_c} \\
  \dfrac{\partial V_2}{\partial S_a} & \dfrac{\partial V_2}{\partial S_b} & \dfrac{\partial V_2}{\partial S_c} \\
\end{array}  \right]
\end{equation}

Note that the terms corresponding to (5) in $\mathcal{J}$ are real and defined in terms of real and reactive power demand ($P_T$, $Q_T$), and absolute values of phase voltages ($|V_D|$). On the other hand, the terms corresponding to (6) in $\mathcal{J}$ are defined in terms of complex quantities i.e. complex power demand $S_T$ and complex sequence voltages $V_T$. Due to the different modalities of terms in $\mathcal{J}$, (5) and (6) cannot be solved simply using the update rule specified in (13). To solve this problem first, we write the update equations for solving (5) and (6) using $\mathcal{J}$ separately in (15). Next, we develop an iterative Newton-based method to solve for $\Delta S_T$ and $\Delta V_D$ by iteratively solving (15) using $\mathcal{J}$ defined in (14).
\begin{equation}
\small
 \nonumber  \left[\begin{array}{c}
    \Delta  P_T \\
     \Delta Q_T
    \end{array}\right]
    - \dfrac{\partial f_2(V_D(n))}{\partial V_D(n)}  \left[\begin{array}{c}
      |\Delta V_D| \\
      |\Delta V_D|
    \end{array}\right] =
    \left[\begin{array}{c}
      P_T(n) - P_D(n) \\
      Q_T(n) - Q_D(n)
    \end{array}\right]
\end{equation}
\begin{equation}
\small
      -\dfrac{\partial f_1(S_T(n))}{\partial S_T(n)}
    \Delta  S_T + \mathcal{T} \Delta V_D  = \mathcal{T}(V_D(n)) - V_T(n)
\end{equation}

The updates are obtained using the set of equations defined in (16)-(18). Starting with $\Delta S_T=0$, (16) is solved to obtain $\Delta V_D$. Using $\Delta V_D$, (17)-(18) are solved to obtain $\Delta S_T$. The process is repeated until the changes in updates are not significant. Note that the proposed Newton-based method employs exactly the Jacobian operator, $\mathcal{J}$, that is approximated over multiple iterations of (16)-(18). 
\begin{equation}
\small
 \Delta V_D
    = \mathcal{T}^{-1}\left(\mathcal{T}(V_D(n)) - V_T(n)+\dfrac{\partial f_1(S_T(n))}{\partial S_T(n)}
    \Delta  S_T\right)
\end{equation}
\begin{equation}
\small
  \left[\begin{array}{c}
    \Delta  P_T \\
     \Delta Q_T
    \end{array}\right] =
    \left[\begin{array}{c}
      P_T(n) - P_D(n) \\
      Q_T(n) - Q_D(n)
    \end{array}\right]  + \dfrac{\partial f_2(V_D(n))}{\partial V_D(n)}  \left[\begin{array}{c}
      |\Delta V_D| \\
      |\Delta V_D|
    \end{array}\right]
\end{equation}
\begin{equation}
\small
 \Delta  S_T = \Delta P_T + j\Delta Q_T
\end{equation}

\subsubsection{Calculate Jacobian Operator}
We present an approach to derive the expressions for the Jacobian operator, $\mathcal{J}$, using the Thevenin equivalent representations for the two subsystems. The Thevenin equivalent models, as seen from PCC, for both subsystems for a coupled T\&D system are shown in Fig. 2. We define Thevenin equivalent at the PCC for Subsystem-1 using equivalent impedance matrix ($Z_T$) and the voltage at PCC ($V_T$) and for Subsystem-2 using equivalent impedance matrix ($Z_D$) and the voltage at PCC ($V_D$).

\begin{figure}[t]
        \centering    \includegraphics[width=0.4\textwidth]{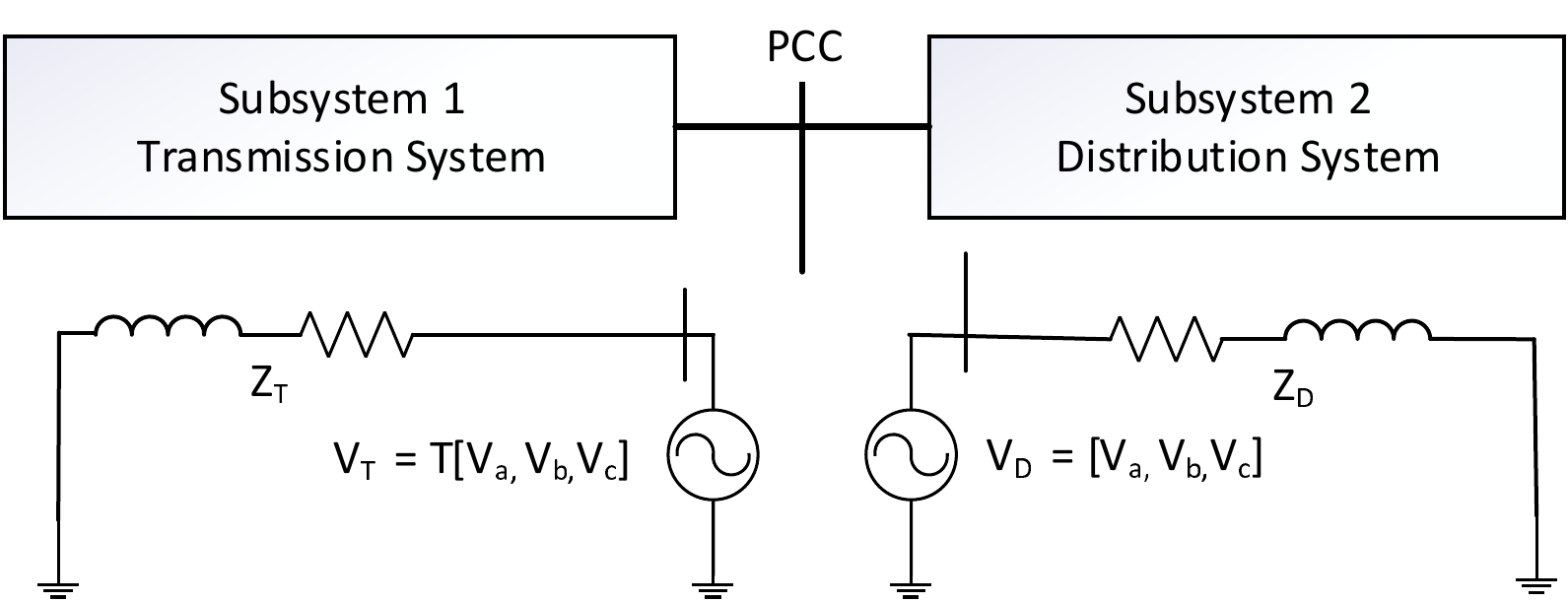}
\vspace{-0.2cm}
        \caption{Thevenin Equivalent for both subsystems.}
        \label{fig2}
\vspace{-0.9cm}
\end{figure}

\vspace{-0.2cm}
\begin{small}
\begin{eqnarray}
\nonumber Z_T=\left[\begin{array}{ccc}
         Z_{aa,T} & Z_{ab,T} & Z_{ac,T} \\
         Z_{ab,T} & Z_{bb,T} & Z_{bc,T} \\
         Z_{ac,T} & Z_{bc,T} & Z_{cc,T}
         \end{array} \right]=\left[\begin{array}{ccc}
         Y_{aa,T} & Y_{ab,T} & Y_{ac,T} \\
         Y_{ab,T} & Y_{bb,T} & Y_{bc,T} \\
         Y_{ac,T} & Y_{bc,T} & Y_{cc,T}
         \end{array} \right]^{-1}
\end{eqnarray}
\end{small}
\vspace{-0.6 cm}

\begin{small}
\begin{eqnarray}
\small
\nonumber Z_D=\left[\begin{array}{ccc}
         Z_{aa,D} & Z_{ab,D} & Z_{ac,D} \\
         Z_{ab,D} & Z_{bb,D} & Z_{bc,D} \\
         Z_{ac,D} & Z_{bc,D} & Z_{cc,D}
         \end{array} \right]         =\left[\begin{array}{ccc}
         Y_{aa,D} & Y_{ab,D} & Y_{ac,D} \\
         Y_{ab,D} & Y_{bb,D} & Y_{bc,D} \\
         Y_{ac,D} & Y_{bc,D} & Y_{cc,D}
         \end{array} \right]^{-1}
\end{eqnarray}
\end{small}

Fist, the analytical expressions for the individual terms in matrix $\dfrac{\partial f_2(V_D)}{\partial V_D}$ are derived using the Thevenin equivalent circuit for Subsystem-2. Note that this matrix represents the change in complex power consumption in Subsystem-2, $S_D$, wrt. the change in input voltages, $V_D$, at the PCC. Using analytical expressions for phase currents and power flows, the individual terms for $\dfrac{\partial f_2(V_D)}{\partial V_D}$ are calculated as a function of $Z_D$ in (19)-(21). The Thevenin equivalent for Subsystem-2 ($Z_D$) is obtained using OpenDSS in fault study mode which is then used to calculate the required matrix using (19)-(21).


\vspace{-0.3cm}
\begin{small}
\begin{eqnarray}
 \nonumber I_a &=& Y_{aa,D}V_a + Y_{ab,D}V_b + Y_{ac,D}V_c\\
  \nonumber S_a &=& P_a + j Q_a = V_a (I_a)^* \\
  \nonumber &=& Y_{aa,D}^*|V_a|^2 + V_a Y_{ab,D}^* V_b^* + V_a Y_{ac,D}^* V_c^*\\
\nonumber \dfrac{\partial S_a}{\partial |V_a|} &=& \dfrac{\partial P_a}{\partial |V_a|} + j\dfrac{\partial Q_a}{\partial |V_a|} \\
\nonumber &=& 2*Y_{aa,D}^*|V_a| + \dfrac{V_a Y_{ab,D}^* V_b^*}{|V_a|} + \dfrac{V_a Y_{ac,D}^* V_c^*}{|V_a|}\\
\nonumber \dfrac{\partial S_a}{\partial |V_b|} &=& \dfrac{\partial P_a}{\partial |V_b|} + j \dfrac{\partial Q_a}{\partial |V_b|} = \dfrac{V_a Y_{ab,D}^* V_b^*}{|V_b|} \\
\dfrac{\partial S_a}{\partial |V_c|} &=& \dfrac{\partial P_a}{\partial |V_c|} + j\dfrac{\partial Q_a}{\partial |V_c|} = \dfrac{V_a Y_{ac,D}^* V_c^*}{|V_c|}
\end{eqnarray}
\end{small}

\vspace{-1cm}

\begin{small}
\begin{eqnarray}
\nonumber I_b &=& Y_{ab,D}V_a + Y_{bb,D}V_b + Y_{bc,D}V_c \\
\nonumber S_b &=& P_b + j Q_b = V_b (I_b)^* \\
  \nonumber &=& V_b Y_{ab,D}^* V_a^* + Y_{bb,D}^* |V_b|^2 + V_b Y_{bc,D}^* V_c^*\\
  \nonumber \dfrac{\partial S_b}{\partial |V_b|} &=& \dfrac{\partial P_b}{\partial |V_b|} + j \dfrac{\partial Q_b}{\partial |V_b|} \\
  \nonumber &=& 2*Y_{bb,D}^*|V_b| + \dfrac{V_b Y_{ab,D}^* V_a^*}{|V_b|} + \dfrac{V_b Y_{bc,D}^* V_c^*}{|V_b|}\\
\nonumber \dfrac{\partial S_b}{\partial |V_a|} &=& \dfrac{\partial P_b}{\partial |V_a|} + j\dfrac{\partial Q_b}{\partial |V_a|} = \dfrac{V_b Y_{ab,D}^* V_a^*}{|V_a|} \\
\dfrac{\partial S_b}{\partial |V_c|} &=& \dfrac{\partial P_b}{\partial |V_c|} + j\dfrac{\partial Q_b}{\partial |V_c|} = \dfrac{V_b Y_{bc,D}^* V_c^*}{|V_c|}
\end{eqnarray}
\end{small}

\vspace{-0.7cm}

\begin{small}
\begin{eqnarray}
 \nonumber I_c &=& Y_{ac,D}V_a + Y_{bc,D}V_b + Y_{cc,D}V_c\\
  \nonumber S_c &=& P_c + j Q_c = V_c (I_c)^* \\
  \nonumber &=& V_c Y_{ac,D}^*V_a^* + V_c Y_{bc,D}^* V_b^* + Y_{cc,D}^* |V_c|^2\\
\nonumber \dfrac{\partial S_c}{\partial |V_c|} &=& \dfrac{\partial P_c}{\partial |V_c|} + j\dfrac{\partial Q_c}{\partial |V_c|} \\
\nonumber &=& 2*Y_{cc,D}^*|V_c| + \dfrac{V_c Y_{ac,D}^* V_a^*}{|V_c|} + \dfrac{V_c Y_{bc,D}^* V_b^*}{|V_c|}\\
\nonumber \dfrac{\partial S_c}{\partial |V_b|} &=& \dfrac{\partial P_c}{\partial |V_b|} + j\dfrac{\partial Q_c}{\partial |V_b|} = \dfrac{V_c Y_{bc,D}^* V_b^*}{|V_b|} \\
\dfrac{\partial S_c}{\partial |V_a|} &=& \dfrac{\partial P_c}{\partial |V_a|} + j\dfrac{\partial Q_c}{\partial |V_a|} = \dfrac{V_c Y_{ac,D}^* V_a^*}{|V_a|}
\end{eqnarray}
\end{small}

Similarly, we define matrix $\dfrac{\partial f_1(S_T)}{\partial S_T}$ using Thevenin equivalent circuit of Subsystem-1 in (22)-(24). This matrix calculates the change in sequence voltages at the PCC, $V_T$, wrt. complex power demand at the PCC, $S_T$. Note that unlike (19)-(21), the expressions in (22)-(24) do not explicitly depend upon the Thevenin impedance matrix of Subsystem-1, $Z_T$. 
Therefore, we do not calculate $Z_T$. The calculation of $\dfrac{\partial f_1(S_T)}{\partial S_T}$ only requires complex current at the T\&D interface which is directly obtained from the Subsystem-1 solver.

\vspace{-0.3cm}

\begin{small}
\begin{eqnarray}
\nonumber \left[\begin{array}{c}
    V_a \\
    V_b \\
    V_c
  \end{array}\right] &=&  \left[\begin{array}{ccc}
         1 & 1 & 1 \\
         1 & a^2 & a \\
         1 & a & a^2
         \end{array} \right] \left[\begin{array}{c}
    V_0 \\
    V_1 \\
    V_2
  \end{array}\right] \\
 \nonumber \left[\begin{array}{c}
    I_a \\
    I_b \\
    I_c
  \end{array}\right] &=& \left[\begin{array}{ccc}
         Z_{aa,T} & Z_{ab,T} & Z_{ac,T} \\
         Z_{ab,T} & Z_{bb,T} & Z_{bc,T} \\
         Z_{ac,T} & Z_{bc,T} & Z_{cc,T}
         \end{array} \right]^{-1} \left[\begin{array}{c}
    V_a \\
    V_b \\
    V_c
  \end{array}\right]\\
 \nonumber S_a &=& (V_0 + V_1 + V_2)I_a^*\\
 \nonumber S_b &=& (V_0 + a^2 V_1 + a V_2)I_b^*\\
 \nonumber S_c &=& (V_0 + a V_1 + a^2 V_2)I_c^*
\end{eqnarray}
\end{small}
\vspace{-0.5cm}
\begin{small}
\begin{eqnarray}
 \dfrac{\partial V_0}{S_a} = \dfrac{1}{I_a^*}, \dfrac{\partial V_1}{S_a} = \dfrac{1}{I_a^*}, \dfrac{\partial V_2}{S_a} = \dfrac{1}{I_a^*} \\
 \dfrac{\partial V_0}{S_b} = \dfrac{1}{I_b^*}, \dfrac{\partial V_1}{S_b} = \dfrac{1}{a^2 I_b^*}, \dfrac{\partial V_2}{S_b} = \dfrac{1}{a I_b^*} \\
 \dfrac{\partial V_0}{S_c} = \dfrac{1}{I_c^*}, \dfrac{\partial V_1}{S_c} = \dfrac{1}{a I_c^*}, \dfrac{\partial V_2}{S_c} = \dfrac{1}{a^2 I_c^*}
\end{eqnarray}
\end{small}
\vspace{-0.3cm}

\subsubsection{Iterative Newton-based Approach}
For a given iteration of co-simulation, the boundary variables obtained from the subsystem solvers i.e. $V_T$ and $S_D$ are available. The current values of boundary variables are used to calculate $\mathcal{J}$ by evaluating (19)-(24). Next, $\mathcal{J}$ is used to obtain the Newton-based updates defined in (16)-(18). Although updates require iteratively solving (16)-(18), in practice not more than two iterations are required to achieve sufficient accuracy. Using the current values of updates, the inputs to the subsystem solvers, i.e. $S_T$ and $V_D$, are modified and subsystems are solved again in parallel. The second iteration of co-simulation begins and $\mathcal{J}$ is recalculated for the new value of boundary variables obtained from the subsystem solvers. This process is repeated until the residuals defined in (7)-(8) are within permissible error tolerance (see Algorithm 1).

Notice that $\mathcal{J}$ needs to be updated at each iteration of the co-simulation. This requires updating $Z_D$ and evaluation of the expressions in (19)-(24). The method to obtain Thevenin impedance based on short-circuit analysis using OpenDSS is rather fast (see \cite{OpenDSS}). Also, since the calculation of $\mathcal{J}$ requires evaluation of a few simple mathematical expressions (see (19)-(24)) and updates are obtained without inverting $\mathcal{J}$ (see (16)-(18)), the Newton-based approach is fast and scalable.

Although we have derived the update rules using only two subsystems, the proposed approach is general and applicable to a transmission system connected to multiple distribution feeders as demonstrated in the results section. For multiple subsystems, the method requires Thevenin impedance matrix ($Z_D$) for all connected distribution systems at their respective substation buses. Here, $Z_D$ is calculated individually and in parallel by the respective distribution solvers (OpenDSS).



\section{Results and Discussions}
The proposed co-simulation approach is validated using an integrated T\&D test system, with IEEE 9-bus transmission system model and one of the EPRI green circuits, Ckt-24, as distribution system model. The IEEE 9-bus transmission test system includes three generators at buses 1, 2 and 3 and three loads at buses 5, 6 and 8 (see Fig. 3) \cite{TransTestFeeder}. 
EPRI Ckt-24 is based on a real-world feeder and supplies for a total of 3885 customers at 34.5 kV voltage level using two equally loaded feeders \cite{DistTestFeeder}. The peak demand recorded at the substation is 52.1 MW and 11.7 MVAR and the feeder is comprised of 87\% single-phase/two-phase residential loads. A substation transformer is used to connect the 34.5 kV distribution system to the 230 kV transmission bus.
     \begin{figure}[h]
      \centering
  \vspace{-10 pt}
      \includegraphics[width=0.45\textwidth]{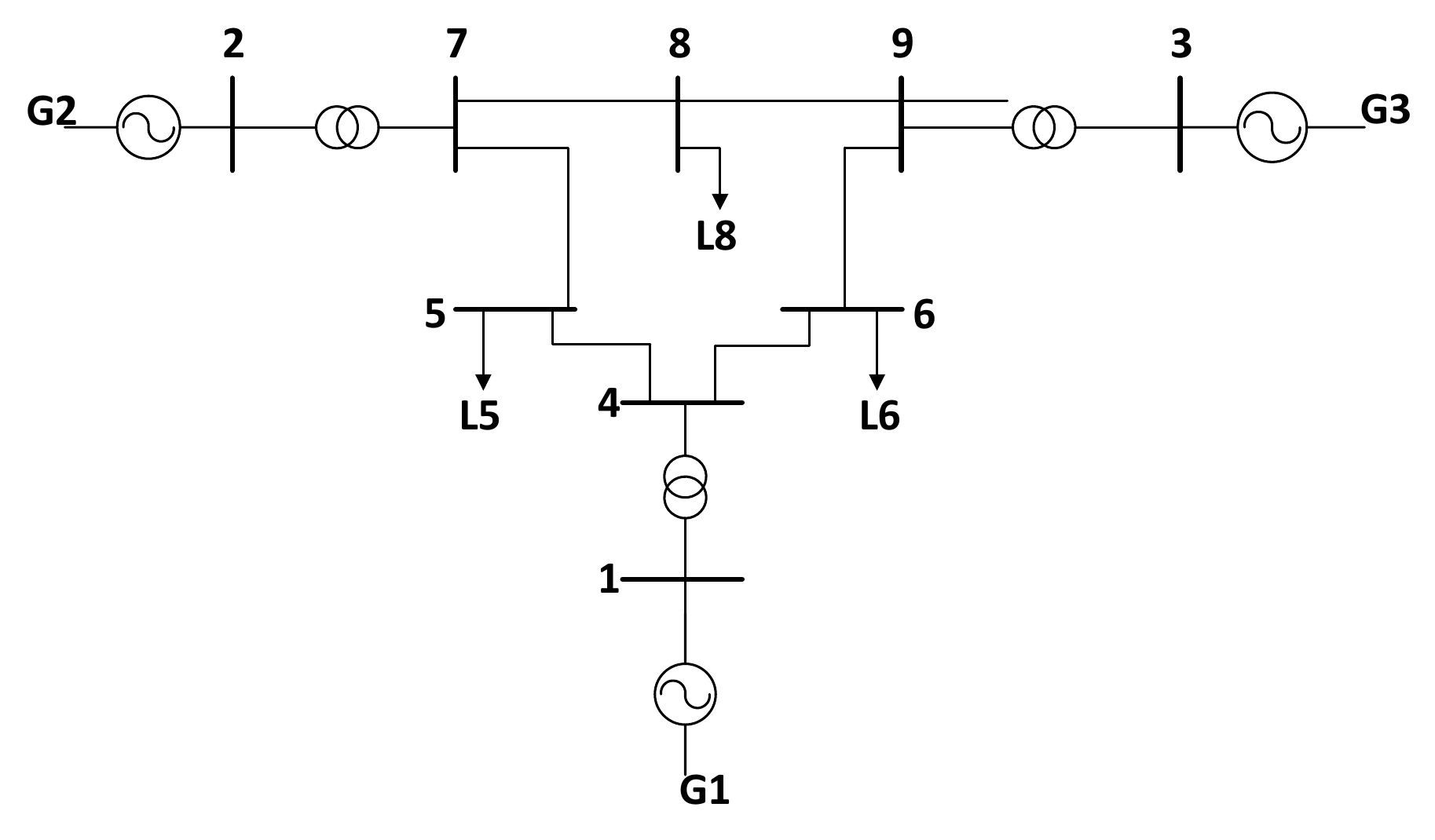}
          \vspace{-6 pt}
      \caption{IEEE 9-bus Transmission Test System \cite{TransTestFeeder}.}
      \label{fig3}
      \vspace{-0.3cm}
  \end{figure}


Two integrated T\&D test systems are developed. Test System-1 (TS1) is obtained by replacing the aggregated load at bus 6 of the 9-bus transmission system with EPRI Ckt-24. Test System-2 (TS2) is obtained by replacing all the load points, L5, L6, and L8, of the IEEE 9-bus system with Ckt-24. 

\vspace{-0.3cm}


\subsection{Convergence for Base Case and Increased System Loading}
The base case load scenario simulates the loading condition with no intentional unbalances in the distribution feeder having an approximate per-phase complex power demand equal to $17.8 MW +5.1j MVar$. The co-simulation approach is implemented for both test systems and the load demand and bus voltages at PCC for each iteration are shown in Table I. Table I also compares the number of iterations required by FPI method and Newton's method to solve the interface equations. It can be seen from the table that phase voltages converge for both co-simulation methods in four iterations. An error tolerance of $10^{-4}$ is used for convergence.

Next, we compare the values of converged positive sequence voltage at Bus 6 obtained using the proposed three-sequence transmission model vs. those obtained using a positive sequence transmission model; the values for positive sequence voltages are 1.0442 p.u. and 1.0441 p.u., respectively. Since the system is balanced, the converged voltages are approximately the same for both transmission system models. The differences in the converged values for the two models are more pronounced with the increase in system unbalance (see Section IV.C).

    \begin{table}[t]
        \centering
        \caption{TS1: Bus voltage convergence at PCC - Bus 6 }
\vspace{-0.3cm}
        \label{table1}
        \begin{tabular}{c|c|c|c|c|c}
            \toprule[0.4 mm]    
            \multicolumn{6}{c}{Voltage Convergence using FPI Method}\\
            \toprule[0.4 mm]
            \hline
            System & Phase & Iter. 1 & Iter. 2  & Iter. 3 & Iter. 4 \\
                    \hline
                    \multirow{3}{5em}{Transmission} & Phase A & 1.0437 & 1.0435 &  1.0437 & 1.0437\\
                    &    Phase B    & 1.0453 & 1.0451 &  1.0454 & 1.0454\\
                    &    Phase C    & 1.0432 & 1.0430 &  1.0433 & 1.0433\\
                    \hline
                    \multirow{3}{4em}{Distribution} & Phase A & 1.05 & 1.0437 & 1.0435 & 1.0437\\
                    &    Phase B    & 1.05  & 1.0453 &  1.0451 & 1.0454\\
                    &    Phase C    & 1.05 & 1.0432 &  1.0430 & 1.0433\\
                    \hline

            \toprule[0.4 mm]    
            \multicolumn{6}{c}{Voltage Convergence using Newton's method}\\
            \toprule[0.4 mm]
            \hline
            System & Phase & Iter. 1 & Iter. 2  & Iter. 3 & Iter. 4\\
                    \hline
                    \multirow{3}{5em}{Transmission} & Phase A & 1.0437 & 1.0441 &  1.0437 &  1.0437\\
                    &    Phase B    & 1.0453& 1.0457 &  1.0454 &  1.0454\\
                    &    Phase C    & 1.0432 & 1.0437 &  1.0433 &  1.0433\\
                    \hline
                    \multirow{3}{4em}{Distribution} & Phase A & 1.05 & 1.0440 & 1.0438 &  1.0437\\
                    &    Phase B    & 1.05  & 1.0455 &  1.0455 &  1.0454\\
                    &    Phase C    & 1.05 & 1.0437 &  1.0434 &  1.0433\\
                    \hline
        \end{tabular}
\vspace{-0.3cm}
    \end{table}

    \begin{table}[t]
    \centering
    \caption{Number of iterations(N) and time(T) required for convergence}
    \vspace{-0.3cm}
    \label{table2}
    \begin{tabular}{c|c|c|c|c|c|c|c|c}    
        \hline
        &\multicolumn{4}{c|}{FPI Method}&\multicolumn{4}{c}{Newton's Method}\\
        \hline
        Load multiplier &\multicolumn{2}{c|}{ TS-1} & \multicolumn{2}{c|}{ TS-2} & \multicolumn{2}{c|}{ TS-1}  & \multicolumn{2}{c}{ TS-2} \\
        \hline
        & N & T (s) & N & T (s) & N & T (s)& N & T (s) \\
        \hline
        $1$ & 4 & 1.83 & 4 & 2.71 & 4 & 3.18 & 6 & 4.71  \\
        \hline
        $1.5$ & 6 & 2.34  & 7 & 5.42 & 5 & 4.77 & 7 & 7.15  \\
        \hline
        $2$ & 8 & 5.84 & 9 & 8.17  & 6 & 5.12  & 7 & 6.51   \\
        \hline
        $2.5$ & 10 & 6.18  & 10 & 9.14 & 7 & 5.74 & 7 & 7.42  \\
        \hline
    \end{tabular}
    \vspace{-1.2 cm}
\end{table}
    
Both test systems are also compared for different loading conditions. The load multiplier for the distribution feeder is adjusted to increase the demand in all phases. This results in an increased loading without causing any additional unbalance. The results are shown in Table II. In general, as the system loading is increased, both FPI and Newton's method take a longer time to converge with a proportional increase in the number of iterations. However, Newton's method converges faster with lesser number of iterations as the system loading increases as shown for cases with load multiplier 2 and 2.5.

\vspace{-0.2cm}    
\subsection{Model Convergence for Load Unbalance}
The proposed co-simulation methods are compared for multiple load unbalance scenarios for both test systems, TS1 and TS2, where unbalance is defined using ANSI C84.1 \cite{ANSI}. The load unbalance is simulated by modifying load allocation factors for the loads connected to one of the phases. We select Phase A to simulate low loading conditions by modifying the load allocation factor. To this regard, both current and voltage unbalance are of interest. Table III shows eight unbalanced loading scenarios simulated for both test systems. The unbalance is characterized by the percentage of current and voltage unbalance. For Cases 1-4, current unbalance seen at the PCC is varied from 20\%-60\%. It is noted that percentage voltage unbalance increases with the increase in percentage current unbalance. Similar to Cases 1-4, Cases 5-8 also simulate a current unbalance of 20\%-60\% but after increasing the loading of distribution system by 1.5 times. An increased level of voltage unbalance is recorded due to the increased loading. 

\begin{table}[t]
    \centering
    \caption{Simulated cases for unbalanced load conditions}
    \vspace{-0.3cm}
    \label{table3}
    \begin{tabular}{c|c|c|c|c|c|c|c|c}    
        \hline
        Unbalance  &\multicolumn{8}{c}{Cases}\\
        \cline{2-9}
        & 1 & 2 & 3& 4& 5& 6& 7& 8 \\
        \hline
        Current(\%)& 20 & 40 & 50& 60& 20& 40& 50& 60 \\
        \hline
        Voltage(\%)& 0.22 & 0.80 & 1.43& 2.00& 0.39& 1.35& 2.37& 3.00 \\
        \hline
        $S_{T}$(MVA)& 45 & 60 & 80 & 95 & 67.5 & 90 & 120 & 142.5 \\
        \hline
    \end{tabular}
    \vspace{-0.4 cm}
\end{table}

\begin{table}[t]
    \centering
    \caption{Number of iterations(N) and time(T) required for convergence for Unbalanced Load Conditions}
    \label{tablew}
    \vspace{-0.3cm}
    \begin{tabular}{c|c|c|c|c|c|c|c|c}    
    \hline
    &\multicolumn{4}{c|}{FPI Method}&\multicolumn{4}{c}{Newton's Method}\\
    \cline{2-9}
    Unbalance &\multicolumn{2}{c|}{ TS-1} & \multicolumn{2}{c|}{ TS-2} & \multicolumn{2}{c|}{ TS-1}  & \multicolumn{2}{c}{ TS-2} \\
    \cline{2-9}
        cases & N & T (s) & N & T (s) & N & T (s)& N & T (s) \\
        \hline
        Case $1$ & 6 & 4.61 &  6  & 6.91 &  5 & 5.83 & 5 & 6.32 \\
        \hline
        Case $2$ & 9 & 6.47 & 10 & 10.71 & 7 & 6.14 & 7 & 7.14 \\
        \hline
        Case $3$ & 11 & 7.73 & 12 & 13.28 & 7 & 6.72 & 8 & 7.51 \\
        \hline
        Case $4$ & 14 & 10.46 & 16 & 17.23 & 9 & 7.51 & 9 & 9.03  \\
        \hline
        Case $5$ & 8 & 8.14 & 9 & 11.81 & 6 & 5.74 & 7 & 6.95 \\
        \hline
        Case $6$ & 12 & 12.06 & 13 & 15.07 & 7 & 8.12 & 8 & 9.71 \\
        \hline
        Case $7$ & 20 & 18.89 & 30 & 21.34 & 9  & 10.56 & 11 & 14.04 \\
        \hline
        Case $8$ & 28 & 22.18 & 42 & 26.98 & 10 & 12.84 & 13 & 15.33 \\
        \hline
    \end{tabular}
    \vspace{-2.2cm}
\end{table}

The maximum number of iterations and time taken for interface equations to converge for both test systems are detailed in Table IV. Both, the unbalance in currents and voltages, affect the system convergence characteristics. For Cases 1-8, as current unbalance is increased, the time and the number of iterations taken to converge by FPI and Newton's method increases. However, Newton's method takes fewer iterations to converge in significantly lesser time. The degree of voltage unbalance also affects the time taken for the system to converge. For instance, both Case 4 and Case 8, result in 60\% current unbalance, but due to increased loading in Case 8, the voltage unbalance is higher. Due to this, the FPI method takes 42 iterations to converge in 26.98 sec for Case 8 as compared to 16 iterations in 17.23 sec for Case 4 when solving Test System-2 (TS2). The performance of Newton's method is significantly better than the FPI method and the improvements are more pronounced for higher levels of system unbalance. In fact, for Case 8, Newton's method takes 13 iterations in  15.53 sec as compared to 42 iterations in 26.98 sec taken by FPI method for TS2. 
This demonstrates the efficiency and scalability of  Newton's method in highly stressed environments. Another key observation is that the FPI method takes a higher number of iterations for single vs. multiple feeder case. The performance of Newton's method, however, is approximately the same for both single and multi-feeder cases. This would be of significant advantage of using Newton's method for the analysis of a large-scale integrated T\&D system with multiple interconnected distribution feeders. 

\begin{figure}[t]
    \centering
    \vspace{-0.3cm}
    \includegraphics[trim={0.1cm 0 0 0cm},clip,width=0.45\textwidth]{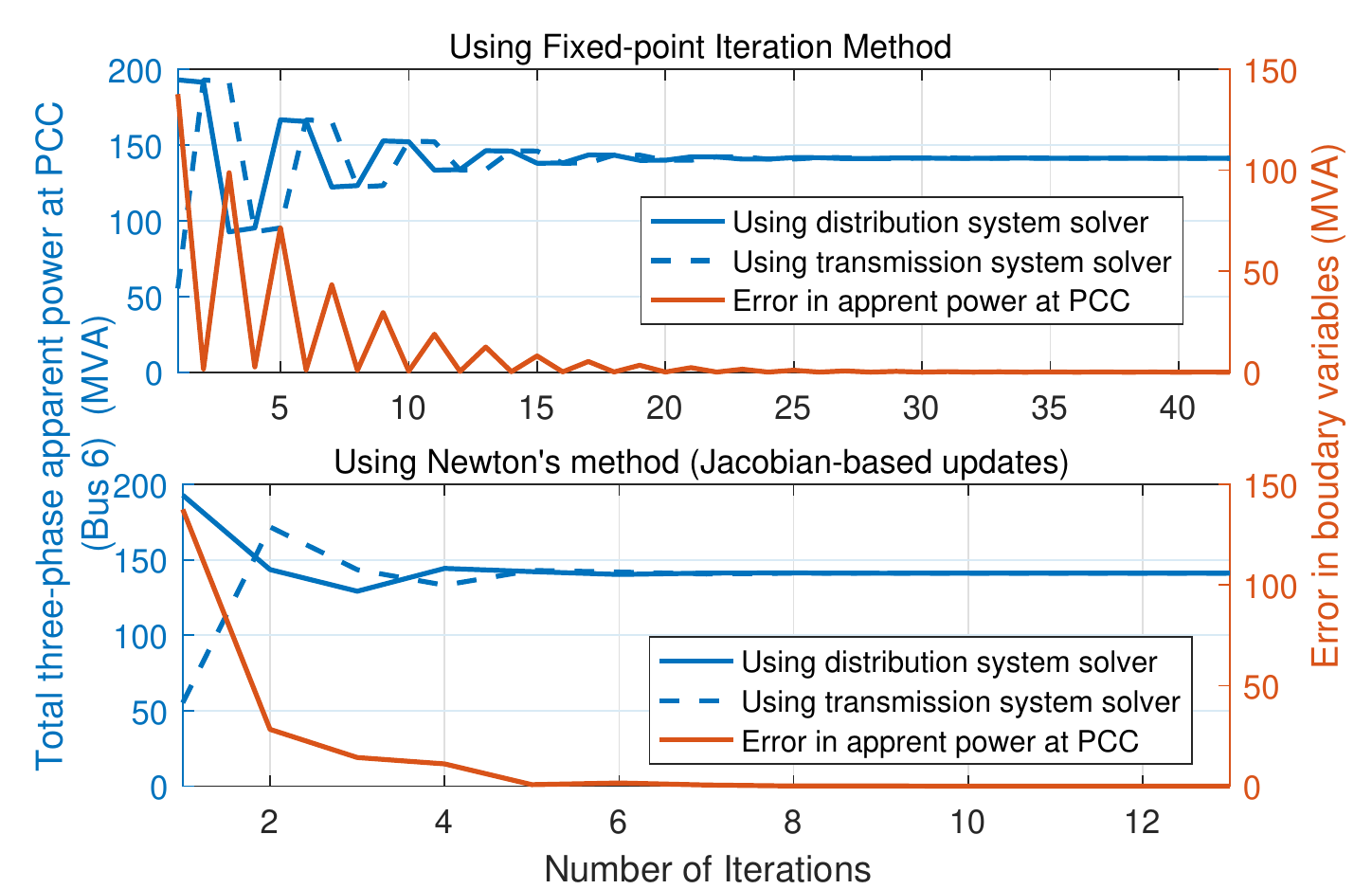}
    \vspace{-0.4cm}
    \caption{Convergence of total three-phase apparent power demand at bus-6 for TS2 with unbalanced load condition corresponding to Case 8.}
    \label{fig:7}
    \vspace{-0.2cm}
\end{figure}

\begin{figure}[t]
    \centering
    \vspace{-0.2cm}
    \includegraphics[trim={0.5cm 0 0 0cm},clip,width=0.45\textwidth]{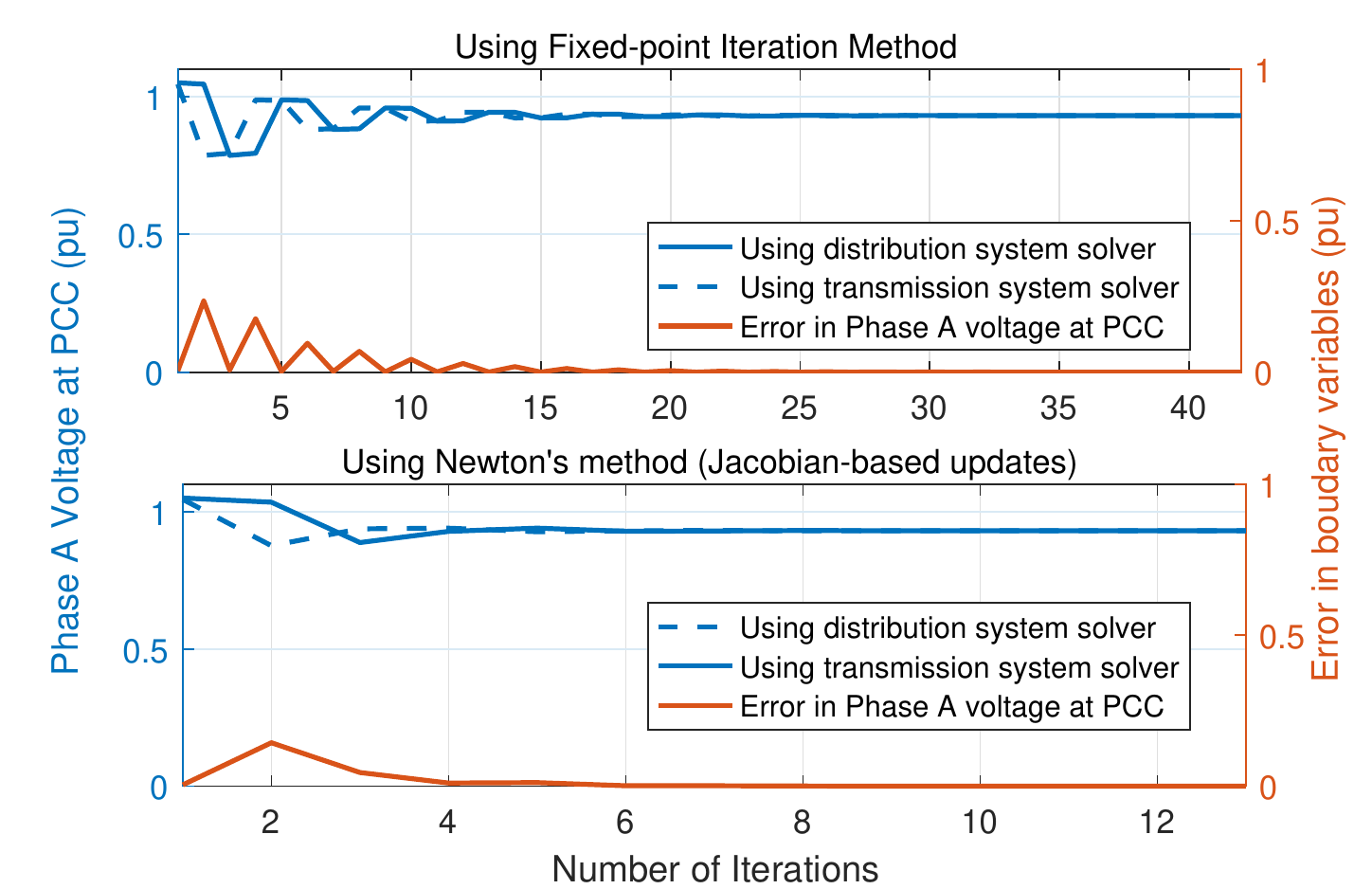}
    \vspace{-0.4cm}
    \caption{Convergence of Phase-A voltage magnitude at bus-6 for TS2 with unbalanced load condition corresponding to Case 8}
    \label{fig:8}
    \vspace{-0.9cm}
\end{figure}

To further describe the solution methodology for FPI and Newton's method, the solution for interface equations for each iteration corresponding to Case 8 for TS2 are analyzed. The plot for total apparent power at PCC obtained at each iteration using distribution system and transmission system solvers for the two methods are shown in Fig. 4. Similarly, plots for voltage at Phase A of the PCC obtained using distribution and transmission system solvers using two methods are shown in Fig. 5. Note that while solutions and errors for FPI method oscillate before converging, the error for Newton's method decreases at each iteration, thus allowing for faster convergence.

\vspace{-0.2cm}
\subsection{Three-Sequence vs. Positive-Sequence Transmission Model}
This section compares the results for co-simulation obtained using a three-sequence transmission model vs. a positive-sequence transmission model. The results for the converged positive sequence voltages at Bus-6 for the two cases are shown in Table V for different levels of system unbalance. As the unbalance increases, the difference in the converged positive sequence voltages obtained using the two models increases significantly. This indicates that during unbalanced load conditions, using a positive-sequence transmission system model may lead to significant errors in T\&D co-simulation.
\begin{table}[t]
    \centering
    \caption{Comparison of converged positive sequence voltage at PCC (Three sequence vs. positive sequence transmission model)}
    \label{comptable}
    \begin{tabular}{c|c|c}    
        \hline
        Unbalance  & With three-sequence  & With positive sequence \\
        cases & transmission model & transmission model \\
        \cline{2-3}
        & V (p.u.) & V (p.u.) \\
        \hline
        Case $1$ & 1.0382 &    1.0319  \\
        \hline
        Case $2$ &1.0208 &    1.0164  \\
        \hline
        Case $3$ & 0.9913 &    0.9767  \\
        \hline
        Case $4$ & 0.9719 &    0.9489  \\
        \hline
        Case $5$ & 1.0067 &    1.0003 \\
        \hline
        Case $6$ &     0.9721 & 0.9635  \\
        \hline
        Case $7$ & 0.9371 &    0.9129 \\
        \hline
        Case $8$ & 0.9247 &    0.9018 \\
        \hline
    \end{tabular}
\end{table}

To further validate the results, in our prior work, a similar system was simulated in a standalone environment and the results obtained using the proposed co-simulation approach (with three-sequence transmission model) were compared against those obtained from the standalone model \cite{DubeyNAPS2018}. It was validated that the proposed approach closely approximates the standalone model, especially during unbalanced conditions.
\vspace{-0.2cm}

\subsection{Iteratively Coupled vs. Loosely Coupled Method}
In this section, the loosely coupled method for co-simulation approach is compared against the proposed iteratively coupled co-simulation model. The test case includes a time-series T\&D analysis with high load variability. This case is possible when the distribution feeder is supplying for a high penetration of DERs especially PVs incurring high generation variability. Here, the simulation is carried out for a 60-min duration, where T\&D systems are solved at every 1-min interval. 

The total three-phase apparent power demand measured at the PCC is shown in Fig. 6. It can be seen that the apparent power demand obtained using the loosely coupled model do not match with those obtained using the iteratively coupled co-simulation approach. The differences are higher during time-stamps when load changes are sudden and drastic. Since, the proposed iteratively coupled model allows for convergence of boundary variables at every time-stamp, the obtained solutions represent the actual results for the integrated T\&D system analysis (also see \cite{DubeyNAPS2018}). The errors incurred in loosely coupled model indicate that it is not able to converge to actual values of system variables for the cases with high load variability.

 \begin{figure}[t]
      \centering
    \vspace{-0.2cm}
      \includegraphics[trim={12 0 12 12},clip,width=0.46\textwidth]{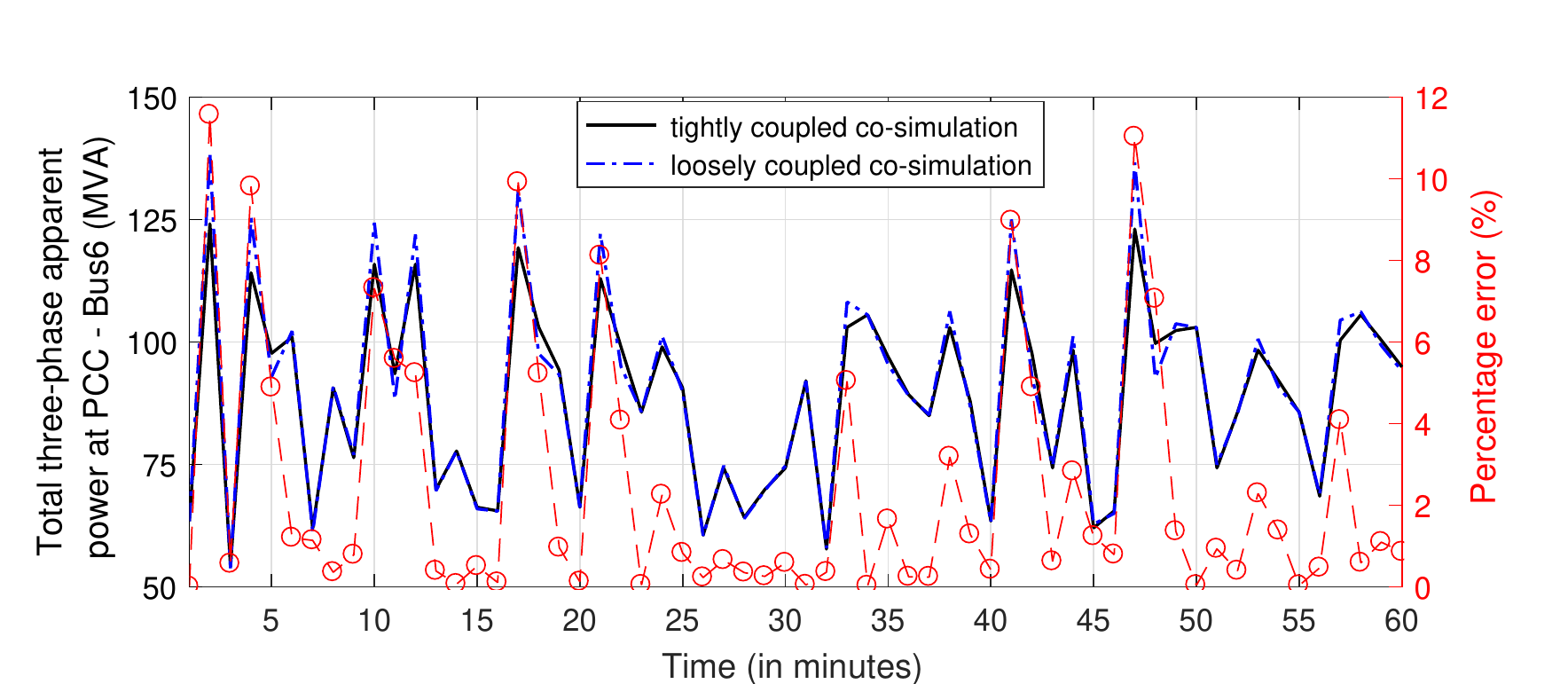}
    \vspace{-0.3cm}
      \caption{Comparing iteratively coupled and loosely coupled methods.}
      \label{fig:11}
      \vspace{-0.8cm}
  \end{figure}

\vspace{-0.2cm}
\subsection{Co-simulation models vs Standalone-model}
{\color{blue}{  
The co-simulation methods developed in this work are compared against a monolithic solver for the stand-alone T\&D model. For stand-alone simulation, IEEE 9-bus system with three-phase details and three Ckt-24 distribution systems connected to its load nodes are simulated in OpenDSS. The power flow equations for the simulated stand-alone T\&D model are solved using OpenDSS. Originally, in this work, the three distribution systems connected to individual transmission load points were solved by a series implementation (SI) technique. This means that following each solution of transmission system, the three distribution systems are solved independently but is a sequence (one after the other). Please refer to Sections II.B and II.C for more details on the SI technique. Typically, all three distribution systems can be solved together and in parallel by distributing the individual distribution systems to different computer cores \cite{palmintier2016experiences}. We call this approach the parallel implementation (PI) of co-simulation method. To test PI co-simulation method for added computational advantage, we have implemented both FPI and Newton-based co-simulation approach using the MATLAB Parallel Computing Toolbox. In this implementation, the three distribution systems connected to IEEE 9-bus system are solved simultaneously i.e. in parallel.

The time taken for solving the eight test cases described in Table III by co-simulation methods (both SI and PI implementation) are compared against the stand-alone solver. The results are shown in Fig. 7. The Newton's method with parallel implementation converges the fastest for all test cases. Generally, all co-simulation methods converge relatively faster than the stand-alone model except in Case 8 when FPI method in SI mode takes longer time to converge. However, for the same case (case 8), the convergence time improves significantly on implementing the FPI method in PI mode i.e. when the three distribution systems are solved in parallel. Overall, the co-simulation models converge faster than the stand-alone model for the selected test system especially with parallel implementation. } }

  \begin{figure}[t]
  \includegraphics[width=0.46\textwidth]{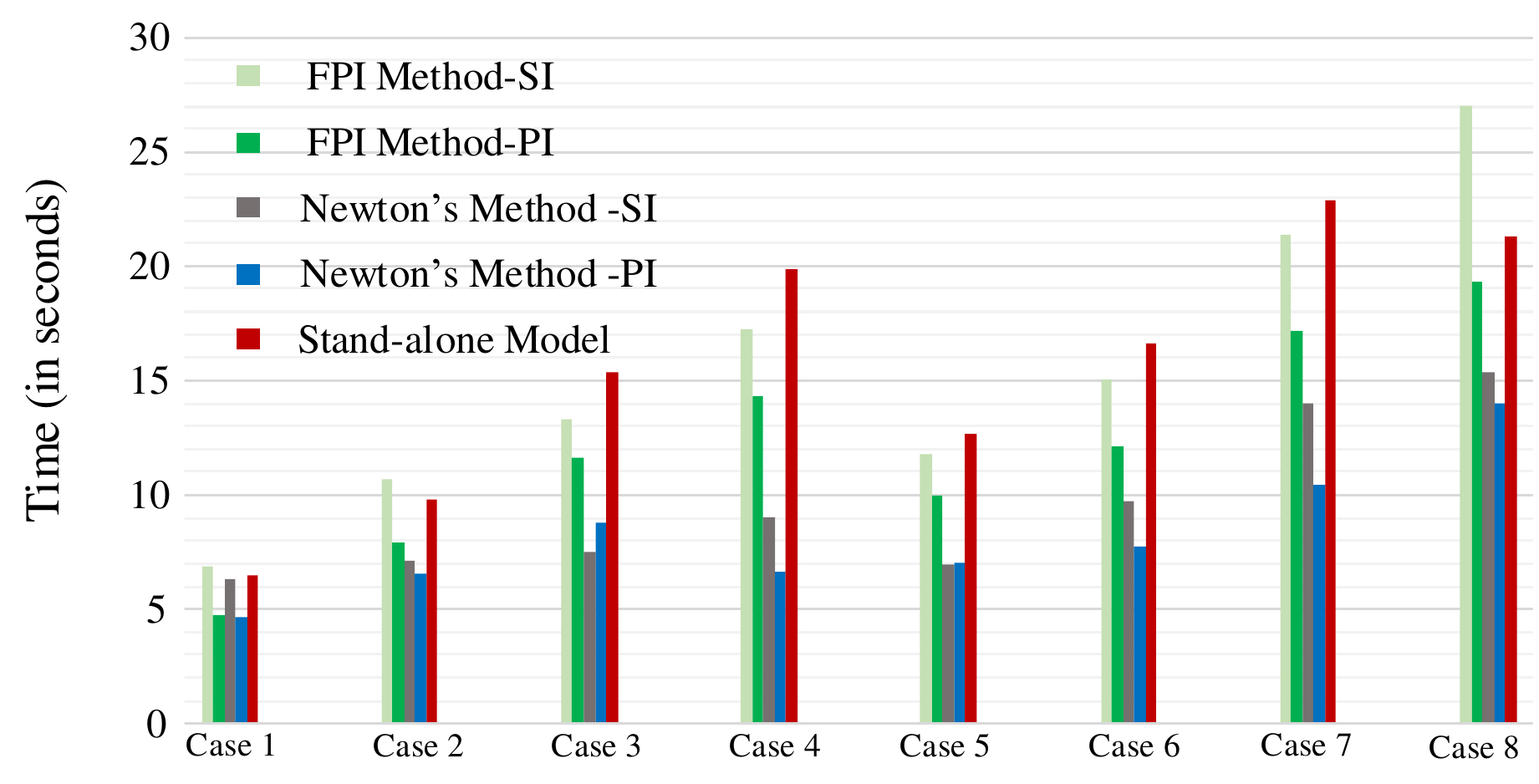}
     \caption{Comparison of time taken for convergence in Test System-2 using co-simulation model and stand-alone model}
     \label{fig:12}
   \vspace{-0.9cm}
   \end{figure}

\vspace{-0.25cm}
\section{Conclusion}
This paper presents an iteratively coupled co-simulation framework for an unbalanced integrated T\&D system analysis. The primary objective is to bring co-simulation close to a standalone T\&D system model that can accurately model unbalanced load conditions and increased demand variability likely to be realized in feeders with an increased level of DER penetrations. The proposed framework is comprised of a three-sequence AC power flow for the transmission system, a three-phase AC power flow for the distribution system, and an iterative coupling approach for T\&D interface. An analytical framework is developed to mathematically represent the T\&D co-simulation interface and first-order and second-order convergent methods, using FPI and Newton's method, respectively, are proposed to solve the nonlinear interface equations. The results conclude that the proposed framework converges for different levels of system loading and unbalance conditions. Newton's method requires less time to converge as compared to FPI method and the stand-alone model. The improvements in the number of iterations and the time taken to converge are more pronounced for stressed system conditions.
    
    \bibliographystyle{IEEEtran}
    \vspace{-0.2cm}
    \bibliography{IEEEabrv,references}
    
    

\end{document}